\documentclass[11pt,superscriptaddress,aps,prd,preprint]{revtex4}

\everymath{\displaystyle}
\usepackage{graphicx}

\usepackage[T1]{fontenc}
\usepackage{amsmath}
\usepackage{amssymb}
\usepackage{graphicx}

\newcommand{\bea}{\begin{eqnarray}}
\newcommand{\eea}{\end{eqnarray}}

\begin{document}


\title{On Lorentz violation in $e^{-}\!\!+\!e^{+}\!\rightarrow\!\mu^{-}\!\!+\!\mu^{+}$ scattering at finite temperature}

\author{P. R. A. Souza}
\email{pablo@fisica.ufmt.br}
\affiliation{Instituto de F\'{\i}sica, Universidade Federal de Mato Grosso,\\
78060-900, Cuiab\'{a}, Mato Grosso, Brazil}

\author{A. F. Santos}
\email{alesandroferreira@fisica.ufmt.br}
\affiliation{Instituto de F\'{\i}sica, Universidade Federal de Mato Grosso,\\
78060-900, Cuiab\'{a}, Mato Grosso, Brazil}

\author{S. C. Ulhoa}
\email{sc.ulhoa@gmail.com}
\affiliation{International Center of Physics, Instituto de F\'isica, Universidade de Bras\'ilia, 70910-900, Bras\'ilia, DF, Brazil}

\author{F. C. Khanna \footnote{Professor Emeritus - Physics Department, Theoretical Physics Institute, University of Alberta\\
Edmonton, Alberta, Canada}}
\email{khannaf@uvic.ca}
\affiliation{Department of Physics and Astronomy, University of
Victoria,BC V8P 5C2, Canada}

\begin{abstract}

Small violation of Lorentz and CPT symmetries may emerge in models unifying gravity with other forces of nature. An extension of the standard model with all possible terms that violate Lorentz and CPT symmetries are included. Here a CPT-even non-minimal coupling term is added to the covariant derivative. This leads to a new interaction term that breaks the Lorentz symmetry. Our main objective is to calculate the cross section for the  $e^{-}\!\!+\!e^{+}\!\rightarrow\!\mu^{-}\!\!+\!\mu^{+}$  scattering  in order to investigate any violation of Lorentz and/or CPT symmetry at finite temperature. Thermo Field Dynamics formalism is used to consider finite temperature effects.

\end{abstract}

\maketitle

\section{Introduction} 

The Standard Model (SM) is a successful field theory that describes fundamental particles and their interactions with high precision. The SM is a gauge theory with the symmetry group $\mathrm{U(1)\times SU(2)\times SU(3)}$ \cite{1,2}. The SM is a fundamental theory. However it does not include the theory of gravitation in its framework, that includes three fundamental forces of nature: the electromagnetic, weak and strong forces. There are several attempts to unify all interactions of nature in a unique fundamental theory. Among various candidates for a unified theory the most famous is the string theory \cite{14}. In addition this model does not explain in a satisfactory way some problems such as, the hierarchy problem  \cite{6}, the neutrino oscillation \cite{7}, cosmic particles at high energies  \cite{8, 10}, among others. Including these issues leads to a physics well beyond the standard model.

It is anticipated that a fundamental theory would emerge at very high energies ($\approx10^{19}\,\mathrm{GeV}$). At sufficiently high energies, the possibility of small violation of the Lorentz and CPT symmetries may be present. Some models, such as string theory \cite{Kostelecky1}, lead to spontaneous breaking of Lorentz symmetry. It is interesting to note that a quantum theory of gravitation may be anticipated to violate  the Lorentz symmetry. These ideas lead to the Standard Model Extension (SME) that violates Lorentz and CPT symmetry. Such models have been developed \cite{Kostelecky2, Kostelecky3, Kostelecky4}. The SME consists of models of well known physics of the SM plus all possible terms that violate Lorentz and CPT symmetry. In addition, it is divided into two parts: (i) the minimal version restricted to power counting renormalizable operators and (ii) the non-minimal version which also includes operators of higher dimensions.

The structure of SME is a way to investigate the Lorentz violation. However, there is another interesting way to investigate the Lorentz violation that modifies the interaction between fermions and photons, i.e., a new non-minimal coupling term added to the covariant derivative \cite{Belich1}. The non-minimal coupling term may be CPT-odd or CPT-even that have been considered for various applications  \cite{Belich1, Belich2, Belich3, Belich4, Belich5, SP, Brito, Casana1, Casana2, Casana3, Casana4}. Here a CPT-even non-minimal coupling term will be included to analyze the  $e^{-}\!\!+\!e^{+}\!\rightarrow\!\mu^{-}\!\!+\!\mu^{+}$ scattering, a well-known quantum electrodynamics process, at finite temperature. The Thermo Field Dynamics (TFD) formalism will be used to introduce temperature effects.

TFD is a real time finite temperature formalism \cite{Umezawa1, Umezawa2, Umezawa22, Khanna1, Khanna2, Kbook}. It includes the statistical average of an observable ${\cal A}$ expressed as a thermal vacuum expectation value i.e., $\langle {\cal A} \rangle=\langle 0(\beta)| {\cal A}|0(\beta) \rangle$, where $|0(\beta) \rangle$ is the thermal vacuum, $\beta=\frac{1}{k_BT}$, with $T$ being the temperature and $k_B$ is the Boltzmann constant (we use $k_B=\hbar=c=1$). This formalism is composed of two ingredients, the doubling of the Hilbert space and the Bogoliubov transformation. This doubling consists of the Hilbert space composed of the original space, $S$, and a fictitious space (tilde space), $\tilde{S}$. The map between the tilde and non-tilde operators is defined by the tilde (or dual) conjugation rules. The temperature effect is implemented in the doubled Hilbert space by a Bogoliubov transformation which introduces a rotation of the tilde and non-tilde variables.

This paper is organized as follows. In section II, an introduction to the TFD formalism is developed. In the section III, the model is presented. The transition amplitude and the cross section for three different vertices are calculated. In section IV, some concluding remarks are presented.

\section{TFD formalism}

TFD is a thermal quantum field theory with a thermal vacuum $|0(\beta)\rangle$. It is composed by two fundamental ingredients: (1) doubling the degrees of freedom in a  Hilbert space and (2) the Bogoliubov transformation. The expanded Hilbert space is defined as $S_T=S\otimes \tilde{S} $, with $S$ being the standard Hilbert space and $\tilde{S}$ the fictitious Hilbert space. The map between the tilde $\tilde{B_i}$ and non-tilde $B_i$ operators is defined by the following tilde conjugation rules:
\bea
(B_iB_j)^\thicksim = \tilde{B_i}\tilde{B_j}, \quad\quad (cB_i+B_j)^\thicksim = c^*\tilde{B_i}+\tilde{B_j}, \quad\quad (B_i^\dagger)^\thicksim = \tilde{B_i}^\dagger, \quad\quad (\tilde{B_i})^\thicksim = -\xi B_i,
\eea
with $\xi = -1$ for bosons and $\xi = +1$ for fermions. The Bogoliubov transformation introduces a rotation in the tilde and non-tilde Hilbert space with thermal dependence. The Bogoliubov transformation is different for fermions and bosons. Here our interest is in fermions. Then the Bogoliubov transformation for fermions is
\begin{eqnarray}\label{2.3}
c_p&=&{\mathtt u}(\beta)c_p(\beta)+{\mathtt v}(\beta)\tilde{c}^\dagger_p(\beta),\nonumber\\
c^\dagger_p&=&{\mathtt u}(\beta)c^\dagger_p(\beta)+{\mathtt v}(\beta)\tilde{c}_p(\beta),\\
\tilde{c}_p&=&{\mathtt u}(\beta)\tilde{c}_p(\beta)-{\mathtt v}(\beta)c^\dagger_p(\beta),\nonumber\\
\tilde{c}^\dagger_p&=&{\mathtt u}(\beta)\tilde{c}^\dagger_p(\beta)-{\mathtt v}(\beta)c_p(\beta),\nonumber\
\end{eqnarray}
where $c_p$ and $c^\dagger_p$ are the annihilation and creation operators. The factors $\mathtt{u}(\beta)$ and $\mathtt{v}(\beta)$ are given as
\begin{eqnarray}\label{2.4}
&&\mathtt{u}(\beta)=\cos(\theta(\beta))=(e^{-\beta|\kappa_0|}+1)^{-1},\\
&&\mathtt{v}(\beta)=\sin(\theta(\beta))=(e^{\beta|\kappa_0|}+1)^{-1}.\nonumber\
\end{eqnarray}
Algebraic rules for thermal operators are
\begin{eqnarray}\label{2.5}
&&\{ c_{p}(\beta), c^\dagger_{q}(\beta) \}=\delta^3(p-q),\\
&&\{ \tilde{c}_{p}(\beta), \tilde{c}^\dagger_{q}(\beta) \}=\delta^3(p-q),
\end{eqnarray}
and other anti-commutation relations are null.

In the framework of the TFD formalism the transition amplitude for any QED process is given as
\begin{equation}\label{2.7}
\hat{\mathcal{S}}_{fi}(\beta)=\Big\langle f,\beta\Big|\hat{S}\Big|i,\beta\Big\rangle,
\end{equation}
where the thermal states are defined as
\begin{eqnarray}\label{2.8}
&&|i,\beta\rangle=c^\dagger_{p_1}(\beta,s_1)d^\dagger_{p_2}(\beta,s_2)|0(\beta)\rangle,\\
&&|f,\beta\rangle=c^\dagger_{p_3}(\beta,s_3)d^\dagger_{p_4}(\beta,s_4)|0(\beta)\rangle,
\end{eqnarray}
with $s_i$ being the spin variable ($i=1,2,3,4$) and $\hat{S}$-matrix is defined as
\begin{equation}\label{2.9}
\hat{S}=\sum_{n=0}^{\infty}\frac{(-\imath)^n}{n!}\int dx_1dx_2...dx_n\!:\![\hat{H}_I(x_1)\hat{H}_I(x_2)...\hat{H}_I(x_n)]\!:\,,
\end{equation}
where $\mathrm{\hat{H}_I(x)}=\mathrm{H_I(x)-\tilde{H}_I(x)}$ is the interaction hamiltonian. Here up to the second order term is considered and has the form 
\begin{equation}\label{2.10}
  \hat{S}^{(2)}=\frac{(-\imath)^2}{2}\int d^4xd^4y\!:\![\hat{H}_I(x_1)\hat{H}_I(x_2)]:=S^{(2)}-\tilde{S}^{(2)}.
\end{equation}
Then the transition amplitude becomes
\begin{equation}\label{2.12}
\mathcal{S}_{fi}(\beta)=\Big\langle f,\beta\Big|{S}^{(2)}|i,\beta\Big\rangle=\frac{(-\imath)^2}{2!}\!\!\int\!\!d^4xd^4y\Big\langle f,\beta\Big|\!:\![\mathcal{L}_I(x)\mathcal{L}_I(y)]\!:\!\Big|i,\beta\Big\rangle.
\end{equation}
It is important to note that, there is a similar equation for the tilde part. As the physical quantities are given by non-tilde part, only this part is considered.

Using the transition amplitude, the cross section for any scattering process at finite temperature is considered. The cross section is defined as
\begin{equation}\label{2.13}
\frac{d\sigma(\beta)}{d\Omega}= \frac{1}{64\pi^2}\frac{1}{4s}\sum_{\mathrm{Spin}}\Big|\mathcal{S}_{fi}(\beta)\Big|^2,
\end{equation}
where $\sqrt{s}\!=\!2E=E_{_{CM}}$ and $E_{_{CM}}$ is the center of mass (CM) energy. In addition an average over the spin of the incoming particles and summing over the spin of outgoing particles is included.

In the next section the transition amplitude will be calculated. Then the cross section for the $e^{-}\!\!+\!e^{+}\!\rightarrow\!\mu^{-}\!\!+\!\mu^{+}$ scattering at finite temperature is calculated.

\section{Cross section of the $e^{-}+e^{+}\longrightarrow\mu^{-}+\mu^{+}$ scattering}

Here the cross section for the $e^{-}+e^{+}\longrightarrow\mu^{-}+\mu^{+}$ scattering at finite temperature is calculated. In addition Lorentz-violating effects are included. The Lorentz violation is using a non-minimal coupling term that is added to the covariant derivative, i.e.,
\begin{equation}\label{3.1}
\mathfrak{D}_\mu=D_\mu+\frac{\lambda}{2}K_{\mu\nu\theta\rho}\gamma^\nu F^{\theta\rho},
\end{equation}
with $D_\mu=\partial_\mu+i eA_\mu$. Here $\lambda$  is the coupling constant for Lorentz violation term. The tensor $K_{\mu\nu\theta\rho}$ belongs to the CPT-even gauge sector of the SME. It has the same symmetries as that of the Riemann tensor and it possesses double null trace. Thus the interaction part of the Dirac Lagrangian becomes
\begin{equation}\label{3.4}
\mathcal{L}_{_{\mathrm{D}}}^{I}=-e\overline{\Psi}\gamma^\mu\Psi A_\mu+\frac{\lambda}{2}K_{\mu\nu\theta\rho}\overline{\Psi}\Sigma^{\mu\nu}\Psi F^{\theta\rho},
\end{equation}
where $\Sigma_{\mu\nu}=\frac{i}{2}[\gamma_\mu\,,\gamma_\nu]$ is used. The first term describes the usual QED vertex and the second term is a new vertex that implies violation of Lorentz symmetries due to the CPT-even tensor. The tensor $K_{\mu\nu\theta\rho}$ may be decomposed into birefringent and non-birefringent
components. Here our investigation is restricted to the non-birefringent components that is represented by a symmetric and traceless rank-2 tensor $K_{\mu\nu}$ \cite{Kostnon}, i.e.,
\begin{equation}\label{3.6}
K_{\mu\nu\alpha\beta}=\frac{1}{2}\Big[g_{\mu\alpha}K_{\sigma\beta}-g_{\nu\alpha}K_{\mu\beta}+g_{\nu\beta}K_{\mu\alpha}-g_{\mu\beta}K_{\nu\alpha}\Big],
\end{equation}
where $K_{\mu\nu}$ is defined by the contraction $K_{\mu\nu}\equiv K^\rho\,_{\mu\rho\nu}$. Then the interaction Lagrangian becomes
\begin{equation}\label{3.8}
\mathcal{L}_{_{\mathrm{D}}}^{I}=-e\overline{\Psi}\gamma^\mu\Psi A_\mu+\lambda\overline{\Psi}\Big(\Sigma_{\beta\nu} K^{\nu\mu}-\Sigma^{\mu\nu} K_{\nu\beta}\Big)\kappa^\beta A_\mu\Psi,
\end{equation}
with $\kappa_\mu$ being the 4-momentum of the photon. This interaction Lagrangian implies the following vertices:
\begin{eqnarray}
 &&\bullet\rightarrow\mathrm{V}^\mu_{(0)}=-i e\gamma^\mu,\label{3.9}\\
 &&\otimes\rightarrow\mathrm{V}^\mu_{(1)}=-i\lambda\kappa^\beta\Big(\Sigma_{\beta\nu} K^{\nu\mu}-\Sigma^{\mu\nu} K_{\nu\beta}\Big)\label{3.10}.
\end{eqnarray}
The Feynman diagrams that describe this scattering process are given in FIG.1.
\begin{figure}[h]
\centering
\includegraphics[scale=0.4]{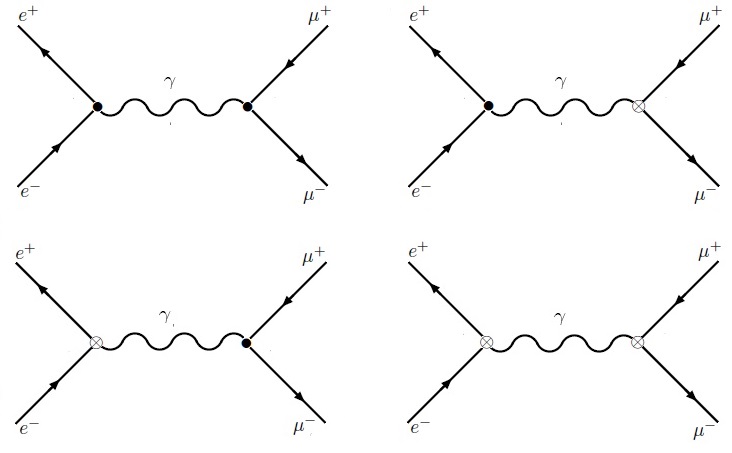}
 \caption{Tree-level Feynman diagrams with different vertices}
\end{figure}

To analyze this process, consider the center of mass frame (CM) such that
\begin{eqnarray}\label{3.11}
&&p_1=(E,p^i),\quad\quad\quad p_3=(E,p'^{i}),\nonumber\\
&&p_2=(E,-p^i),\quad\quad p_4=(E,-p'^{i}),\nonumber\\
&&\kappa=(p_1+p_2)=(\sqrt{s},0),\nonumber\\
\end{eqnarray}
with $p_1,\,p_2,\,p_3$ and $p_4$ being the 4-momentum of the electron, positron, muon and anti-muon, respectively. The new vertex components are
\bea
\mathrm{V}_{(1)}^0&=&0,\\
\mathrm{V}_{(1)}^i&=&\mathrm{V}_{\!\!\mathrm{+is}}^i+\mathrm{V}_{\!\!\mathrm{+an}}^i+\mathrm{V}_\mathrm{-}^i
\eea
where the part associated with the parity-even isotropic coefficient is
\begin{eqnarray}
\mathrm{V}_{\!\!\mathrm{+is}}^\mathrm{i}=-\imath\sqrt{s}K_{00}\Sigma^\mathrm{0i},\label{3.12}
\end{eqnarray}
the anisotropic parity-even part is
\begin{eqnarray}
\mathrm{V}_{\!\!\mathrm{+an}}^\mathrm{i}=\imath\sqrt{s}K^\mathrm{ij}\Sigma_\mathrm{0j},\label{3.13}
\end{eqnarray}
and the parity-odd component is
\begin{eqnarray}
\mathrm{V}_{\!\mathrm{-}}^\mathrm{i}=-\imath\sqrt{s}K_\mathrm{j}\Sigma^\mathrm{ij}.\label{3.14}
\end{eqnarray}
Then the transition amplitude is written as
\begin{equation}\label{3.15}
 \mathcal{S}_{\!fi\,\lambda}(\beta)=\frac{1}{2}\,\mathrm{\int\!d^4xd^4y}\sum_\mathrm{a,b}\Big\langle f,\beta\Big|\!:\overline{\Psi}(\mathrm{x})\mathrm{V}_{(a)}^\mu\Psi(\mathrm{x})\overline{\Psi}(\mathrm{y})\mathrm{V}_{(b)}^\nu\Psi(\mathrm{y}) A_\mu(x)A_\nu(y):\!\Big|i, \beta\Big\rangle,
\end{equation}
with $a, b = 0, 1$. Considering that the wave function of the fermion field is
\begin{equation}\label{3.3}
\Psi(\mathrm{x})=\int\!\!d\mathrm{p}\Big[c_\mathrm{p}(s)u(\mathrm{p},s)e^{-\imath \mathrm{px}}+d_\mathrm{p}^\dagger(s) v(\mathrm{p},s)e^{\imath \mathrm{px}}\Big],
\end{equation}
with $c_p$ and $d_p$ being annihilation operators for electrons and positrons, respectively with $u(p,s)$ and $v(p,s)$ being Dirac spinors, then eq. (\ref{3.15}) becomes
\bea
  \mathcal{S}_{fi\lambda}(\beta)&=&\!\!\int\!\frac{\mathrm{d}^4p}{(2\pi)^4}\mathrm{\int d^4xd^4y}e^{-\imath\mathrm{x}(p_1-p_3)-\imath\mathrm{y}(p_2-p_4)}\nonumber\\
&\times&\sum_\mathrm{a,b}\Big[\overline{v}(p_2,s_2)\mathrm{V}_{(a)}^\mu u(p_1,s_1)\Big]\Big[\overline{u}(p_3,s_3)\mathrm{V}_{(b)}^\nu v(p_4,s_4)\Big]\nonumber\\
&\times&\Big\langle 0(\beta)\Big|\!:\!A_\mu(\mathrm{x})A_\nu(\mathrm{y})\!:\!\Big|0(\beta)\Big\rangle,
\eea
where the Bogoliubov transformation and the anti-commutation relation between the annihilation and creation operators have been used. The photon propagator at finite temperature \cite{Umezawa2, Kbook, our} is given as
\begin{equation}\label{2.14}
\Big\langle 0(\beta)\Big|\!:\!A_\mu(\mathrm{x})A_\nu(\mathrm{y})\!:\!\Big|0(\beta)\Big\rangle=i\!\!\int\!\!\frac{d^4\kappa}{(2\pi)^4}e^{-i\kappa\mathrm{(x-y)}}
\Big[\Delta^{\!f}_0(\kappa)-\Delta^{\!f}_\beta(\kappa)\Big]\eta_{\mu\nu},
\end{equation}
where
\begin{equation}\label{2.15}
  \Delta^{\!{f}}_0(\kappa)=\frac{1}{\kappa^2}\left(
                                       \begin{array}{cc}
                                         1 & 0 \\
                                         0 & -1 \\
                                       \end{array}
                                     \right),
\end{equation}
is the zero temperature part of the photon propagator and
\begin{equation}\label{2.16}
\Delta^{\!{f}}_\beta(\kappa)=\frac{2\pi i\delta(\kappa^2)}{e^{\beta|\kappa_0|}-1}\left(
                                       \begin{array}{cc}
                                         1 & e^{\beta|\kappa_0|/2} \\
                                         e^{\beta|\kappa_0|/2} & -1 \\
                                       \end{array}
                                     \right),
\end{equation}
is the finite temperature part. Using the definition of the four-dimensional delta function and carrying out the $\kappa$ integral, the matrix element becomes
\begin{equation}\label{3.18}
 \mathcal{S}_{\!fi\,\lambda}(\beta)=i\Big(\mathtt{u}^2(\beta)-\mathtt{v}^2(\beta)\Big)^2\Big[\Delta^f_0(\kappa)-
 \Delta^f_\beta(\kappa)\Big]\mathcal{S}_{\!fi\,\lambda},
\end{equation}
with
\begin{equation}\label{3.19}
\mathcal{S}_{\!fi\,\lambda}=\frac{1}{\kappa^2}\sum_{a,b=0}^1\Big[\overline{v}(p_2,s_2)V_{(a)}^\mu u(p_1,s_1)\Big]\Big[\overline{u}(p_3,s_3)V_{(b)\mu} v(p_4,s_4)\Big],
\end{equation}
being the matrix element at zero temperature. The remaining delta function that expresses overall four-momentum conservation is ignored. Using the relation
\bea
[\overline{v}_2\mathrm{V}^\mu_au_1][\overline{u}_1\mathrm{V}_{b\mu}v_2]=\mathrm{tr}[\mathrm{V}^\mu_au_1\overline{u}_1\mathrm{V}^\mu_bv_2\overline{v}_2]
\eea
and the eq. (\ref{2.4}) for the functions $\mathtt{u}(\beta)$ and $\mathtt{v}(\beta)$, the square of the transition amplitude is found as
\begin{eqnarray}\label{3.20}
\sum_\mathrm{spin}|\mathcal{S}_{\!fi\,\lambda}(\beta)|^2&=&\frac{\mathcal{B}(\beta)}{s^2}\sum_\mathrm{a,b}
\sum_\mathrm{c,d}\mathbb{E}^{\mu\nu}_\mathrm{(a,b)}\mathbb{M}_\mathrm{(c,d)\,{\mu\nu}}\,,\nonumber
\end{eqnarray}
where
\begin{equation}\label{3.21}
\mathcal{B}(\beta)=\tanh^4\!\left(\frac{\beta\mathrm{E}_{_{CM}}}{2}\right)\!\!
 \bigg[1+\frac{(2\pi)^2\delta^2(s)}{(e^{\beta\mathrm{E}_{_{CM}}}-1)^2}\bigg].
\end{equation}
Here only the physical component of the photon propagator is considered
\begin{eqnarray}\label{3.22}
  \mathbb{E}^{\mu\nu}_{(a,b)}\hspace{-.3cm}&=&\!\!\!\mathrm{V}_\mathrm{(a)}^{\mu}\sum_{s_1}u(p_1,s_1)
  \overline{u}(p_1,s_1)\overline{\mathrm{V}}_\mathrm{(b)}^{\nu}\sum_{s_2}v(p_2,s_2)
  \overline{v}(p_2,s_2)\nonumber\\
  \!\!\!\!\!&=&\!\!\!\mathrm{tr}\Big[\mathrm{V}_\mathrm{(a)}^{\mu}(p_1\hspace{-.33cm}\slash\,+m_e\,)
  \overline{\mathrm{V}}_\mathrm{(b)}^{\nu}(p_2\hspace{-.33cm}\slash\,-m_e\,) \Big],\\
  \mathbb{M}^{\mu\nu}_{(a,b)}\hspace{-.3cm}&=&\!\!\!\mathrm{V}_\mathrm{(a)}^{\mu}\sum_{s_3}u(p_3,s_3)
  \overline{u}(p_3,s_3)\overline{\mathrm{V}}_\mathrm{(b)}^{\nu}\sum_{s_4}v(p_3,s_4)
  \overline{v}(p_4,s_4)\nonumber\\
 \!\!\!\!\!&=&\!\!\!\mathrm{tr}\Big[\mathrm{V}_\mathrm{(c),\mu}(p_3\hspace{-.32cm}\slash\,+m_\mu\,)
 \overline{\mathrm{V}}_\mathrm{(d)\,\nu}(p_4\hspace{-.32cm}\slash\,-m_\mu\,)\Big],\
\end{eqnarray}
where the relations,
\bea
\sum_\mathrm{s}u(p,s)\overline{u}(p,s)&=&p\hspace{-.16cm}\slash+m\\
\sum_\mathrm{s}v(p,s)\overline{v}(p,s)&=&p\hspace{-.16cm}\slash-m,
\eea
are used. The propagator at finite temperature introduces product of delta functions with identical arguments (\ref{3.21}). This problem is avoided by working with the regularized form of delta-functions and their derivatives \cite{Van}:
\bea
2\pi i\delta^n(x)=\left(-\frac{1}{x+i\epsilon}\right)^{n+1}-\left(-\frac{1}{x-i\epsilon}\right)^{n+1}.
\eea

Thus the differential cross section at finite temperature for this scattering is 
\begin{equation}\label{3.23}
  \mathrm{\frac{d\sigma_\lambda(\beta)}{d\Omega}=\mathcal{B}(\beta)\frac{d\sigma_\lambda}{d\Omega}},
\end{equation}
where
\begin{equation}\label{3.24}
  \mathrm{\frac{d\sigma_\lambda}{d\Omega}=}\frac{1}{(8\pi)^24s^3}\!\sum_\mathrm{a,b}
\sum_\mathrm{c,d}\mathbb{E}^{\mu\nu}_\mathrm{(a,b)}\mathbb{M}_\mathrm{(c,d)\,{\mu\nu}},
\end{equation}
is the differential cross section at zero temperature. Then the cross section at finite temperature  has the form
\begin{equation}\label{3.25}
\sigma_\lambda(\beta)=\mathcal{B}(\beta)\sigma_\lambda,
\end{equation}
with
\begin{equation}\label{3.26}
\sigma_\lambda=\frac{1}{64\pi^2}\frac{1}{4s^3}\,\mathrm{\sum_\mathrm{a,b}\sum_\mathrm{c,d}\mathbb{E}_{(a,b)}^{\mu\nu}\!
\int\!\!\mathrm{d\Omega}\,\mathbb{M}_{(c,d)\mu\nu}},
\end{equation}
where the integration is only on angular variables of scattered particles.

Now let us consider the contribution of each vertex given in eqs. (\ref{3.12}), (\ref{3.13}) and (\ref{3.14}) in the ultra-relativistic limit. In this limit assume $m_e=m_\mu=0$, then the electronic and muonic contributions become
\begin{eqnarray}
  &&\mathbb{E}^{\mu\nu}_\mathrm{(a,b)}\!=\!\mathrm{tr}\Big[\mathrm{V}_\mathrm{(a)}^{\mu}\,p_1\hspace{-.33cm}\slash\,
  \,\overline{\mathrm{V}}_\mathrm{(b)}^{\nu}\,p_2\hspace{-.33cm}\slash\,\,\Big],\label{4.1}\\
  &&\mathbb{M}^{\mu\nu}_\mathrm{(a,b)}\!=\!\mathrm{tr}\Big[\,\mathrm{V}^\mu_\mathrm{(c)}\,p_3\hspace{-.327cm}\slash
 \,\,\overline{\mathrm{V}}^\nu_\mathrm{(d)}\,p_4\hspace{-.33cm}\slash\,\,\Big].\label{4.2}\
\end{eqnarray}
It is important to note that, the non-null components are those with $\mathrm{a\!=\!b}$. When $\mathrm{a\!\neq\!b}$ there are an odd number of Dirac matrices and their trace is zero. They are also null when $\mu\!=\!0$ or $\nu\!=\!0$, since this implies that $\mathrm{V^0_{(1)}}\!=\!0$ and then, $\mathbb{E}_{\mathrm{(ab)}}^{0i}\!\!=\mathbb{E}_{\mathrm{(ab)}}^{i0}$ $\!\!=\mathbb{M}_{\mathrm{(ab)}}^{0i}\!\!=\mathbb{M}_{\mathrm{(ab)}}^{i0}\!\!=0$. Therefore the non-zero components are
\begin{eqnarray}
  &&\mathbb{E}_{(0,0)}^{\mathrm{\,ij}}=\mathrm{tr}[\,\mathrm{V}^\mathrm{i}_{(0)}p_1\!\!\!\!\!\slash\,\,\,
  \mathrm{V}^\mathrm{j}_{(0)}p_2\!\!\!\!\!\slash\,\,\,],\label{4.3} \\
  &&\mathbb{E}_{(1,1)}^{\mathrm{\,ij}}=\mathrm{tr}[\,\mathrm{V}^{\mathrm{i}}_{(1)}p_1\!\!\!\!\!\slash\,\,
  \mathrm{V}^\mathrm{j}_{(1)}p_2\!\!\!\!\!\slash\,\,\,],\label{4.4} \\
  &&\mathbb{M}_{(0,0)}^{\mathrm{\,ij}}=\mathrm{tr}[\,\mathrm{V}^\mathrm{i}_{(0)}p_4\!\!\!\!\!\slash\,\,\,
  \mathrm{V}^\mathrm{j}_{(0)}p_3\!\!\!\!\!\slash\,\,\,],\label{4.5} \\
  &&\mathbb{M}_{(1,1)}^{\mathrm{\,ij}}=\mathrm{tr}[\,\mathrm{V}^{\mathrm{i}}_{(1)}p_4\!\!\!\!\!\slash\,\,
  \mathrm{V}^\mathrm{j}_{(1)}p_3\!\!\!\!\!\slash\,\,\,].\label{4.6}
\end{eqnarray}
Using these results the cross section becomes
\bea\label{4.7}
\sigma_\lambda&=&\frac{1}{64\pi^2}\frac{1}{4s^3}\,
\Bigl(
\mathbb{E}_{(0,0)}^{\,ij}\!\int\!\!d\Omega\,\mathbb{M}_{(0,0)\,ij}+
\mathbb{E}_{(0,0)}^{\,ij}\!\int\!\!d\Omega\,\mathbb{M}_{(1,1)\,ij}\nonumber\\
&+&\mathbb{E}_{(1,1)}^{\,ij}\!\int\!\!d\Omega\,\mathbb{M}_{(0,0)\,ij}+\mathbb{E}_{(1,1)}^{\,ij}\!\int\!d\Omega\,\mathbb{M}_{(1,1)\,ij}
\Bigl).
\eea

\subsection{Isotropic parity-even contribution}

In this case the vertex is given by eq. (\ref{3.12}) and then the non-zero components of eqs. (\ref{4.3})-(\ref{4.6}) are
\begin{eqnarray}
&&\mathbb{E}^\mathrm{\,ij}_{(0,0)}=2e^2(s\delta^\mathrm{\,ij}-4p^\mathrm{i}p^\mathrm{j}),\label{4.9}\\
&&\mathbb{E}^\mathrm{\,ij}_{(1,1)}=8\lambda^2sK^2_{00}p^\mathrm{i}p^\mathrm{j}.\label{4.10}\\
&&\mathbb{M}^\mathrm{\,ij}_{(0,0)}=2e^2(s\delta^\mathrm{\,ij}-4p'^{\mathrm{i}}p'^{\mathrm{j}}),\label{4.11}\\
&&\mathbb{M}^\mathrm{\,ij}_{(1,1)}=8\lambda^2sK^2_{00}p'^{\mathrm{i}}p'^{\mathrm{j}}.\label{4.12}
\end{eqnarray}
Then the integrals in eq. (\ref{4.7}) become
\begin{eqnarray}
&&\mathbb{E}^\mathrm{ij}_{(0,0)}\!\int\!d\Omega\,\mathbb{M}_{(0,0)\,\mathrm{ij}}=\frac{16\pi}{3}4s^2e^4,\label{4.13}\\
&&\mathbb{E}^\mathrm{ij}_{(0,0)}\!\int\!d\Omega\,\mathbb{M}_{(1,1)\,\mathrm{ij}}=
\mathbb{E}^\mathrm{ij}_{(1,1)}\!\int\!d\Omega\,\mathbb{M}_{(0,0)\,\mathrm{ij}}=\frac{16\pi}{3}2s^3e^2\lambda^2K^2_{00}\label{4.14}.\
\end{eqnarray}
Here the term $\mathbb{E}_{(1,1)}^{\,ij}\!\int\!d\Omega\,\mathbb{M}_{(1,1)\,ij}$ is ignored, since it is of the fourth order in Lorentz-violating parameter.
Thus the cross section at finite temperature (up to second order in Lorentz-violating parameter) is 
\begin{equation}\label{4.15}
\sigma_{+\lambda}^\mathrm{i}(\beta)=\mathcal{B}(\beta)\sigma_{\!_\mathrm{QED}}\left[1+\left(\frac{\lambda\sqrt{s}K_{00}}{e}\right)^{\!\!2}\right],
\end{equation}
with $\sigma_{\!_\mathrm{QED}}\!=\!64\pi s^2e^4/3$ and $\mathcal{B}(\beta)$ is defined in eq. (\ref{3.21}). When temperature effects go to zero $\mathcal{B}(\beta)\rightarrow 1$, the result is the same as in \cite{Casana1}.

\subsection{Anisotropic parity-even contribution}

Here the vertex is given in eq. (\ref{3.13}). Then electronic factors are given as
\begin{eqnarray}
  &&\mathbb{E}^{\mathrm{ij}}_{(0,0)}= 2e^2(s\delta^\mathrm{ij}-4p^\mathrm{\,i}p^\mathrm{\,j}),\label{4.16}\\
  &&\mathbb{E}^{\mathrm{ij}}_{(1,1)}=8\lambda^2sK^\mathrm{ik}K^\mathrm{j\,l}p_\mathrm{\,l}p_\mathrm{\,k}\label{4.17},\
\end{eqnarray}
with components of the tensor $\mathbb{M}^\mathrm{\mu\nu}_{(a,b)}$ being same if $p$ is changed to $p'$. Then the integrals in eq. (\ref{4.7}) are 
\begin{eqnarray}
&&\mathbb{E}^\mathrm{ij}_{(0,0)}\!\int\!d\Omega\mathbb{M}_{(0,0)\,\mathrm{ij}}=\frac{16\pi}{3}4s^2e^4,\label{4.18} \\
&&\mathbb{E}^\mathrm{ij}_{(0,0)}\!\int\!d\Omega\mathbb{M}_{(1,1)\,\mathrm{ij}}=\frac{16\pi }{3}s^2e^2\lambda^2\bigg(s|\textbf{K}|^2-4(\,p^\mathrm{i}K_\mathrm{ij}\,)^2\bigg),\label{4.19} \\
&&\mathbb{E}_{(1,1)\,\mathrm{ij}}\!\int\!d\Omega\mathbb{M}_{(0,0)\,\mathrm{ij}}=\frac{16\pi}{3}4e^2s^2\lambda^2(\,p^\mathrm{i}K_\mathrm{ij}\,)^2 \
\end{eqnarray}
where $|\textbf{K}|^2\!=\!K^\mathrm{ij}K_\mathrm{ij}$. Therefore the cross section at finite temperature is
\begin{equation}\label{4.20}
\sigma_{\!\!+\lambda}^\mathrm{a}(\beta)=\mathcal{B}(\beta)\sigma_{_{\!\!QED}}\bigg[1+\left(\frac{\lambda}{e}\right)^{\!\!2}
\bigg(\,\left(\sqrt{s}|\textbf{K}|/2\,\right)^{\!2}+(\,p^\mathrm{i}K_\mathrm{ij}\,)^2\bigg)\bigg].
\end{equation}

\subsection{Parity-odd contribution}

To calculate parity-odd contributions the vertex given in eq. (\ref{3.14}) is considered. Then 
\begin{eqnarray}
&&\mathbb{E}^\mathrm{\,ij}_{(0,0)}=2e^2(s\delta^\mathrm{\,ij}-4p^\mathrm{i}p^\mathrm{j}),\label{4.22}\\
&&\mathbb{E}^\mathrm{\,ij}_{(1,1)}=8\lambda^2s\epsilon^\mathrm{ikl}\epsilon^\mathrm{jm\,n}K_\mathrm{k}K_\mathrm{l}p_\mathrm{m}p_\mathrm{n},\label{4.23}\
\end{eqnarray}
and the muons contributions are obtained in a similar way. The relevant integrals are
\begin{eqnarray}
&&\mathbb{E}^\mathrm{ij}_{(0,0)}\!\int\!d\Omega\,\mathbb{M}_{(0,0)\,\mathrm{ij}}=\frac{16\pi}{3}4s^2e^4,\label{4.24}\\
&&\mathbb{E}^\mathrm{ij}_{(0,0)}\!\int\!d\Omega\,\mathbb{M}_{(1,1)\,\mathrm{ij}}=\frac{16\pi }{3}e^2\lambda^2\bigg(s|\textbf{K}|^2+4(\,\textbf{p}\cdot\textbf{K}\,)^2\bigg), \label{4.25}\\
&&\mathbb{E}_{(1,1)\,\mathrm{ij}}\!\int\!d\Omega\,\mathbb{M}_{(0,0)\,\mathrm{ij}}=\frac{16\pi }{3}e^2\lambda^2\bigg(s|\textbf{K}|^2-8(\,\textbf{p}\cdot\textbf{K}\,)^2\bigg), \label{4.26}\
\end{eqnarray}
with $(\textbf{p}\cdot\textbf{K})=p^\mathrm{i}K_\mathrm{j}$. Then the cross section is
\begin{equation}\label{4.27}
\sigma_\mathrm{-}(\beta)=\mathcal{B}(\beta)\sigma_{_{\!\!QED}}\bigg[1+
\left(\frac{\lambda}{2e}\right)^2\bigg(3s|\textbf{K}|-4(|\textbf{p}||\textbf{K}|\cos(\theta))^2\bigg)\bigg],
\end{equation}
where $\theta$ is the angle between the particle beam and the field $\textbf{K}$. 

The results obtained are general and show that the temperature effects modify the cross section of the scattering process for any chosen vertex. In the limit of zero temperature the standard result for the QED modified by Lorentz-violating parameters are recovered, in all cases. These results also indicate that the temperature effects may improve constraints on Lorentz-violating parameter.

\section{Conclusion}

The SME is a framework to study Lorentz and CPT violation that includes the SM, general relativity and all possible terms that violate the Lorentz and CPT symmetries. Another interesting way is to modify the interaction vertex between fermions and photons, i.e., a new non-minimal coupling term added to the covariant derivative. Here a Lorentz violating CPT-even term is chosen to study the $e^{-}\!\!+\!e^{+}\!\rightarrow\!\mu^{-}\!\!+\!\mu^{+}$ scattering at finite temperature. This new coupling has mass dimension equal to $-1$, which leads to a non-renormalizable theory at power counting. However in the present case this does not pose any problem since our interest is in analyzing the tree-level scattering process. The temperature effects are introduced using the TFD formalism. Three different vertices which introduce the Lorentz violation are considered. Then the cross section at finite temperature is calculated. Our results show that the temperature effects modify the cross section. Then new constraints on Lorentz-violating parameter may be imposed by the temperature effects. In addition astrophysical processes may be studied if the temperature is very high.

\section*{Acknowledgments}

This work by A. F. S. is supported by CNPq project 308611/2017-9.


\begin{thebibliography}{21}
\bibitem{1} T. Kioshita, Quantum eletrodynamics, Advanced Series on Directions in High Energy Physics - Vol. 7(Word Scientific, 1990).
\bibitem{2} M.E. Peskin and D.V. Schroeder, An Introduction to Quantum Field Theory (Westview, 1995).
\bibitem{14} E. Witten, Nucl. Phys. B {\bf 443}, 85 (1995).
\bibitem{6} N. Arkani-Hamed, S. Dimopoulos and G. Dvali, Phys. Lett. B {\bf 429}, 263 (1998).
\bibitem{7} V. Barger, D. Marfatia and K. L. Whisnant, The Physics of Neutrinos, (Princeton University Press. ISBN 0-691-12853-7).
\bibitem{8} L. J. Watson, D. J. Mortlock and A. H. Jaffe, Monthly Notices of the Royal Astronomical Society {\bf 418}, 206 (2011); arXiv:1010.0911
\bibitem{10} W. Bernreuther and M. Suzuki, Rev. Mod. Phys. {\bf 63}, 313 (1991).
\bibitem{Kostelecky1} V. A. Kostelecky and S. Samuel, Phys. Rev. D {\bf 39}, 683 (1989); V. A. Kostelecky and R. Potting, Nucl. Phys. B {\bf 359}, 545 (1991).
\bibitem{Kostelecky2} V. A. Kostelecky and R. Potting, Phys. Rev. D {\bf 51}, 3923 (1995).
\bibitem{Kostelecky3} D. Colladay and V. A. Kostelecky, Phys. Rev. D {\bf 55}, 6760 (1997); Phys. Rev. D {\bf 58}, 116002 (1998).
\bibitem{Kostelecky4} V. A. Kostelecky, Phys. Rev. D {\bf 69}, 105009 (2004).
\bibitem{Belich1} H. Belich, T. Costa-Soares, M. M. Ferreira Jr. and J. A. Helay\"el-Neto, Eur. Phys. J. C {\bf 41}, 421 (2005).
\bibitem{Belich2} H. Belich, T. Costa-Soares, M. M. Ferreira Jr., J. A.Helay\"el-Neto and M. T. D. Orlando, Phys. Lett. B {\bf 639}, 678 (2006).
\bibitem{Belich3} H. Belich, L. P. Colatto, T. Costa-Soares, J. A. Helay\"el-Neto and M. T. D. Orlando, Eur. Phys. J. C {\bf 62}, 425 (2009).
\bibitem{Belich4} H. Belich, T. Costa-Soares, M. M. Ferreira Jr., J. A. Helay\"el-Neto, and F. M. O. Moucherek, Phys. Rev. D {\bf 74}, 065009 (2006).
\bibitem{Belich5} H. Belich, M. M. Ferreira Jr., E. O. Silva, and M. T. D. Orlando, Phys. Rev. D {\bf 83}, 125025 (2011).
\bibitem{SP} B. Charneski, M. Gomes, R. V. Maluf and A. J. da Silva,  Phys. Rev. D {\bf 86}, 045003 (2012).
\bibitem{Brito} G. P. de Brito, J. T. G. Junior, D. Kroff, P. C. Malta and C. Marques, Phys. Rev. D {\bf 94}, 056005 (2016).
\bibitem{Casana1} R. Casana, M. M. Ferreira, R. V. Maluf and F. E. P. dos Santos, Phys. Rev. D {\bf 86}, 125033 (2012).
\bibitem{Casana2} R. Casana, M. M. Ferreira Jr, E. O. Silva, E. Passos and F. E. P. dos Santos, Phys. Rev. D {\bf 87}, 047701 (2013).
\bibitem{Casana3} R. Casana, M. M. Ferreira, Jr., R. V. Maluf and F. E. P. dos Santos, Phys. Lett. B {\bf 726}, 815 (2013).
\bibitem{Casana4} J. B. Araujo, R. Casana and M. M. Ferreira Jr., Phys. Rev. D {\bf 92}, 025049 (2015).
\bibitem{Umezawa1}Y. Takahashi and H. Umezawa, Coll. Phenomena {\bf 2}, 55 (1975); Int. Jour. Mod. Phys. B {\bf 10}, 1755 (1996).
\bibitem{Umezawa2}Y. Takahashi, H. Umezawa and H. Matsumoto, Thermofield Dynamics and Condensed States, North-Holland, Amsterdan, (1982).
\bibitem{Umezawa22} H. Umezawa, Advanced Field Theory: Micro, Macro and Thermal Physics, AIP, New York, (1993).
\bibitem{Khanna1} A. E. Santana and F. C. Khanna, Phys. Lett. A {\bf 203}, 68 (1995).
\bibitem{Khanna2} A. E. Santana, F. C. Khanna, H. Chu, and C. Chang, Ann. Phys. {\bf 249}, 481 (1996).
\bibitem{Kbook}F. C. Khanna, A. P. C. Malbouisson, J. M. C. Malboiusson and A. E. Santana, Themal quantum field theory: Algebraic aspects and applications, World Scientific, Singapore, (2009).
\bibitem{Kostnon} V. A. Kostelecky and M. Mewes, Phys.Rev. D {\bf 80}, 015020 (2009). 
\bibitem{our} A. F. Santos and F. C. Khanna,  Int. J. Mod. Phys. A {\bf 31}, 1650122 (2016). 
\bibitem{Van} Ch.G. van Weert, {\it An Introduction to Real- and Imaginary-time Thermal Field Theory}, Lecture notes on Statistical Field Theory (2001).
\end{thebibliography}
\end{document}